\documentclass[12pt]{article}
\usepackage{amssymb,amscd,amsmath}

\newcommand{\lto}{\longrightarrow}
\newcommand{\E}{\mathcal{E}}
\newcommand{\F}{\mathcal{F}}
\newcommand{\Oh}{\mathcal{O}}
\begin{document}

\title{\bf $K$-homology in algebraic geometry and D-branes}

\author{Eunsang Kim\thanks
{Current address : Department of Applied Mathematics,
Hanyang University, Ansan 425-791, Korea.
{\tt email: eskim@wavelet.hanyang.ac.kr}}\\
{\small \textit {BK21 Mathematical Science Division,}}\\
{\small \textit { Seoul National University,
Seoul 151-747, Korea
}}\\  {\small email: {\texttt{eskim@math.snu.ac.kr}}}}

\maketitle

\begin{abstract} In this article, we investigate the role of
Grothendieck groups
of coherent sheaves in the study of D-branes.
We show how global bound state construction in
topological $K$-theory can be adapted to our context, showing that
D-branes wrapping a subvariety are holomorphically
classified by a relative
$K$-group. By taking the
duality between the relative $K$-groups and the $K$-homologies,
we show that D-brane charge of type
IIB superstrings is properly classified by the $K$-homology.
\end{abstract}

\newpage
\section{Introduction}

In ~\cite{witten98}, E. Witten showed that the D-brane charge is
classified by topological $K$-theory using
some ideas of A. Sen ~\cite{sen98}. A full description of all
states with a D-brane wrapping a closed submanifold of the
spacetime can be made in terms of states of a brane-antibrane pair
on the spacetime and such a brane-antibrane pair with a Tachyon
field naturally defines a $K$-theory class that is trivial at
infinity and hence $K$-theory with compact support. The global
bound state construction depends on the topology of the normal
bundle of the D-brane worldvolume in the spacetime.
In the absence of
the Neveu-Schwarz B-field, a single D-brane can wrap a submanifold
if and only if the normal bundle to
the submanifold has a  $\rm{Spin}^c$ structure, or equivalently
the submanifold is a $\rm{Spin}^c$ manifold.
This follows from the world-sheet global anomalies in the presence
of D-branes studied in ~\cite{freedwitten99}.
The topological
obstruction condition to brane wrapping corresponds to
that of Poincar\'e duality in $K$-theory.
It was argued in
\cite{per06} that the Poincar\'e duality makes it more natural for
D-brane charges to take values in analytic $K$-homology groups,
based on the Poilchinski's basic covariant operational definition
of D-branes~\cite{pol95}. In \cite{szabo01}, it was  argued
that $K$-homology is an appropriate setting for the topological
classification of D-brane charges by an extensive use of
$KK$-theory. Also,
in \cite{AST}, it was asserted that
the stable D-brane configurations can be classified by
topological $K$-homology
groups.

In the same spirit of \cite{witten98}, we study how $K$-homology
can be used to describe type IIB D-branes in the context of
algebraic geometry. In this context, the Grothendieck group of
locally free sheaves corresponds to the topological $K$-theory, as
discussed in \cite{BaDoug84},\cite{sharpe99}. On the other hand,
the Grothendieck group of coherent sheaves is referred as the
$K$-homology since it corresponds to the topological $K$-homology,
\cite{BaDoug84}. The Poincar\'e duality  is the identification of
both Grothendieck groups and it holds for nonsingular projective
algebraic varieties. In \cite{sharpe99}, the $K$-homology has been
studied associated with the derived category of coherent sheaves.
In there, it was asserted that the only physically relevant part
of an object in a derived category of coherent sheaves is its
image in the $K$-homology. However, the physical applications of
$K$-homologies has been limited to the smooth case where the
Poincar\'e duality holds.

In this paper, we will consider the $K$-theoretic
Lefschetz type of duality in algebraic geometry, given in
\cite{BaFulMac79}.
The duality is
the identification of a relative version of the Grothendieck
group of locally free sheaves with a $K$-homology group. We shall
argue that the elements in the relative $K$-group can be naturally
interpreted as a brane-antibrane system with a tachyon field
on a nonsingular variety,
in the absence of D-branes
outside of a closed subvariety. The duality implies that such
brane-antibrane systems can be deformed to a system of branes on a
closed subvariety, which corresponds to the $K$-homology groups.
A novelty of using $K$-homology instead of the
Grothendieck group of locally free sheaves is that we can deal
with coherent and locally free sheaves on the same footing. Also,
it is more suitable when we are dealing with singular varieties,
see \cite{BaDoug84} for example.
Based on this,
we shall
show that D-brane charges of type IIB superstrings are
naturally  classified by $K$-homology
groups for complex projective algebraic varieties including
some singular cases. We discuss how  D-branes wrapping a
nonsingular variety are classified by the relative version of the
Grothendieck group of locally free sheaves which will be referred
as the relative $K$-theory.

Throughout this paper we will work in the absence of
the Neveu-Schwarz B-field. Also, we will not work with
space-filling D-branes.

\section{Grothendieck groups of locally free sheaves}

In this section we shall review the Grothendieck group of locally
free sheaves and the relative $K$-theroy. We show how elements in
the relative $K$-group are interpreted as a brane-antibrane pair.

Let $X$ be a complex projective algebraic variety
(or a compact complex manifold). Denote by $\Oh_X$ the sheaf of
regular functions on $X$ (or the sheaf of holomorphic functions).
The sheaf of
sections of a holomorphic vector bundle of rank $r$ on a complex
manifold
$X$ is
locally free which is locally isomorphic to $\Oh_X^r$.
Conversely, every locally free sheaf of $\Oh_X$-modules is
the sheaf of sections of a holomorphic vector bundle, which is
determined up to isomorphism.

Let $K^0_{\mathsf{alg}}(X)$ be the Grothendieck group of
holomorphic vector bundles on $X$. It is the free abelian group on
the isomorphism classes of holomorphic vector bundles on $X$,
modulo the subgroup generated by elements of the form $E-E'-E''$
whenever there is an exact sequence
\[0\lto E'\lto  E\lto E''\lto 0\]
of holomorphic vector bundles on $X$. Equivalently,
$K^0_{\mathsf{alg}}(X)$ can be considered as the Grothendieck
group of locally free sheaves on $X$ as discussed in \cite{Har}
and
\cite{sharpe99}. The elements in $K^0_{\mathsf{alg}}(X)$ will be
denoted by $[E]$ represented by a holomorphic vector bundle $E$.

We also
introduce the relative $K$-groups of holomorphic vector bundles
following \cite{BaFulMac79}.
Let $X$ be a  complex projective algebraic variety and thus $X$
may have singularities.
Choose a
holomorphic embedding $i:X\hookrightarrow Y$, where $Y$ is a nonsingular
complex manifold.
Consider a complex $E_\bullet$
of holomorphic vector bundles on $Y$
\[E_\bullet : 0\lto E_r\lto E_{r-1}\lto\cdots\lto E_0\lto 0\]
which are exact on $Y-X$. Let $K^0_{\mathsf{alg}}(Y,Y-X)$ be the
Grothendieck group on the isomorphism classes of such complexes of
bundles on $Y$. It is the free abelian group on
such complexes modulo elements $E_\bullet-E'_\bullet-E''_\bullet$
where
\[0\lto E'_\bullet \lto E_\bullet\lto E''_\bullet\lto 0\]
is an exact sequence of complexes. The elements in
$K^0_{\mathsf{alg}}(Y,Y-X)$ will be denoted by $[E_\bullet]$,
represented by a complex $E_\bullet$ of holomorphic vector bundles
on $Y$. If a complex $E_\bullet$ is exact on all of $Y$, then it
defines zero element in $K^0_{\mathsf{alg}}(Y,Y-X)$. In
particular, if $X=Y$, then the relative $K$-group
$K^0_{\mathsf{alg}}(Y,\emptyset)$ is identified with the group
$K^0_{\mathsf{alg}}(Y)$ and the identification is given by
$[E_\bullet]\mapsto \sum_i(-1)^i[E_i]$.

Now,
one can describe the complex
$E_\bullet$, which defines a class in $K^0_{\mathsf{alg}}(Y,Y-X)$,
as brane-antibrane pairs on $Y$. To be more precise,
for a complex $E_\bullet$:
\[0\lto E_r\overset{d_r}\lto E_{r-1}\overset{d_{r-1}}
\lto\cdots\lto E_0\lto 0
\]
of holomorphic vector bundles on $Y$ which is exact on $Y-X$, choose
isomorphisms $E_i\cong \text{Ker }d_i\oplus \text{Ker }d_{i-1}$
on $Y-X$. This gives isomorphisms on $Y-X$
\[
\sigma_1:
E_{\text{ev}}
\lto \sum_i\text{Ker }d_i\]
and
\[
\sigma_2:E_{\text{odd}}\lto \sum_i\text{Ker
}d_i,\]
where $E_{\text{ev}}=\sum_k E_{2k}$ and $E_{\text{odd}}=\sum_k
E_{2k+1}$.
The composition $T=\sigma_1^{-1}\circ\sigma_2$ gives a complex
\[
0\lto E_{\text{odd}}\overset{T}\lto E_{\text{ev}} \lto 0
\]
which is exact on  $Y-X$. The map $T$ is a holomorphic section of
the bundle $E_{\text{odd}}^*\otimes E_{\text{ev}}$ and it has the
adjoint $\bar T$ which is a section of
$E_{\text{odd}}\otimes E_{\text{ev}}^*$ such that $T\bar T T=T$.
In other words, the map $T$ is a tachyon field on the
brane-antibrane pair $(E_{\text{ev}},E_{\text{odd}})$ on $Y$. With
this choice of tachyon field $T$,
the system of the brane-antibrane system is in a
vacuum state on $Y-X$. Thus we are considering a system of
brane-antibrane pair in the absence of D-branes outside of $X$.
We can also interpret the tachyon field as a
holomorphic map $T:Y-X\lto \text{U}(N)$, for some large $N$ and
the whole content of D-brane bound state is captured by such
functions, ({\it cf}. \cite{MoWi99}, \cite{DMW}).
In fact, this brane-antibrane system is
deformed to a system of D-branes on $X$, as we will see in the
next section.

\section{$K$-homology in algebraic geometry and duality}

Let $X$ be a projective algebraic variety. A coherent sheaf $\F$
on $X$ admits a local presentation as an exact sequence
$\Oh_X^p\lto \Oh_X^q\lto\F\lto 0$ and thus it has an exact
sequence of sheaves
\[\E_r\lto\E_{r-1}\lto\cdots\lto\E_0\lto\F\lto 0\]
where the $\E_i$ are locally free sheaves on $X$. Note that the
sequence
\[\E_r\lto\E_{r-1}\lto\cdots\lto\E_0\lto 0\]
is a complex of locally free sheaves on $X$ and the complex
$\E_\bullet$ has the homology $H_0(\E_\bullet)=\F$ and all others
are zero by exactness. A typical example of a non-locally free
sheaf is a skyscraper sheaf which is a coherent sheaf supported at
a point.

The Grothendieck group
of coherent sheaves on $X$
is defined by the free abelian
group on the isomorphism classes of
coherent sheaves on $X$, modulo the subgroup generated by
elements of the form $ \mathcal{F}-\mathcal{F}'-\mathcal{F}''$
whenever there is an exact sequence
\[0\lto \mathcal{F}'\lto\mathcal{F}\lto\mathcal{F}''\lto 0
\]
of coherent sheaves on $X$. As given in \cite{BaDoug84}, the
Grothendieck group is  called the $K$-homology of $X$ because of
its relation with topological $K$-homology. The group is denoted
by $K_0^{\mathsf {alg}}(X)$ and the elements in
$K_0^{\mathsf{alg}}(X)$ will be denoted by $[\F]$ represented by a
coherent sheaf $\F$ on $X$.

If $X$ is nonsingular, two groups $K^0_{\mathsf{alg}}(X)$ and
$K_0^{\mathsf{alg}}(X)$ are isomorphic and is called the
Poincar\'e duality. Following \cite{BaFulMac79},
we study here the more general type of
duality which includes the case of singular varieties.
For
any $[\E_\bullet]\in K^0_{\mathsf{alg}}(Y,Y-X)$,
the homology
sheaves $H_i(\mathcal{E}_\bullet)$ of the complex of locally free
sheaves on $Y$ are coherent sheaves on $Y$ which are supported on
$X$ and thus determine a class $[H_i(\mathcal{E}_\bullet)]$ in
$K_0^{\mathsf {alg}}(X)$. We define the homology map
\[h:K^0_{\mathsf{alg}}(Y,Y-X)\lto K_0^{\mathsf{alg}}(X)\]
by
\begin{equation}
h([\E_\bullet])=\sum_i
(-1)^i[H_i(\mathcal{E}_\bullet)].
\end{equation}
The homology map is an
isomorphism if $Y$ is nonsingular. To see this, we define the
inverse map $h^{-1}$. Let $[\F]\in K_0^{\mathsf{alg}}(X)$,
represented by a coherent sheaf $\F$ on $X$. The sheaf $\F$ on $X$
can be extended to all of $Y$ by zero and defines a coherent sheaf
on $Y$, which will be denoted by $i_*\F$. Since $Y$ is
nonsingular, there is a finite locally free resolution of $i_*\F$:
\[0\lto \mathcal{E}_r\lto \mathcal{E}_{r-1}\lto\cdots\lto
\mathcal{E}_0\lto i_*\mathcal{F}\lto 0.\] Then the complex of
locally free sheaves $\E_\bullet$ on $Y$ is exact on $Y-i(X)$ and
hence defines an element $[\E_\bullet]\in
K^0_{\mathsf{alg}}(Y,Y-X)$. We define a map
$\bar{h}:K_0^{\mathsf{alg}}(X)\lto K^0_{\mathsf{alg}}(Y,Y-X)$ by
\[\bar{h}([\F])=[\E_\bullet].\]
The construction of $\bar h$ does not
depend on the choice of locally free resolution of $i_*\F$,
because any two such resolutions are dominated by the third. Hence
the map $\bar{h}$ is well-defined.
Furthermore, it is straightforward to show that $ \bar{h}=h^{-1}$.
Thus we have
\[h:K^0_{\mathsf{alg}}(Y,Y-X)\cong K_0^{\mathsf{alg}}(X),\]
when $Y$ is nonsingular variety.
In particular, when $X=Y$ and $Y$ is nonsingular, the homology map
is the Poincar\'e duality map and we have
\[\text{PD}:K^0_{\mathsf{alg}}(Y)\cong K_0^{\mathsf{alg}}(Y).\]

In Section 3, we have shown that an element in
$K^0_{\mathsf{alg}}(Y,Y-X)$ can be understood as a brane-antibrane
pair with a tachyon field which is a unitary map outside of $X$.
When $X$ is nonsingular, we have an isomorphism
$h:K^0_{\mathsf{alg}}(Y,Y-X)\cong K^0_{\mathsf{alg}}(X)$, via the
Poincar\'e duality.
This
suggests that a system of brane-antibrane pair supported on $X$
can be
deformed to a system of brane-antibrane wrapped on $X$ since
$K^0_{\mathsf{alg}}(X)=K^0_{\mathsf{alg}}(X,\emptyset)$, in the
absence of $D$-branes on  $Y-X$. Furthermore, since the duality
holds for a singular variety $X$, the system on $Y$ is deformed to
a system of branes on $X$.

\section{The global bound state construction}

In \cite{witten98}, E. Witten showed that D-brane charge of Type
II superstrings are classified by  $K$-theory with compact
support. In the same spirit of \cite{witten98},
we will show the analogous statement using the relative
$K$-theory and $K$-homology.
It has been argued in \cite{witten95} that certain singularities
are allowed in the study of D-branes and those are nonsingular in
string theory. So, one can have more general coherent sheaves as
D-brane bound states, as suggested in \cite{HarMoor}. In this
sense, it is more natural to use $K$-homology rather than the
Grothendieck group of locally free sheaves to classify D-brane
bound states. Also, it is suitable to describe a configuration of
D-branes on a singular variety.

First we describe a simple modification of the Witten's argument,
given in \cite{witten98}. Let
$X$ be a codimension 2, closed subvariety of a nonsingular variety
$Y$.
We will
build a $p$-brane wrapping $X$ from a $p+2$ brane-antibrane pair
on $Y$, where $p+1=\text{dim}_{\mathbb{R}}X$.
One can define a holomorphic line bundle $L$ on $Y$ which
is trivialized on $Y-X$. Let
$\mathcal{L}$ be the associated locally free sheaf on $Y$.
Consider the complex of $\Oh_Y$-modules:
\begin{equation}
0\lto \Oh_Y\overset{T}\lto \mathcal{L}\lto 0
\end{equation}
where $T$ is the zero map on $X$ and is a unitary map on $Y-X$.
The
complex (2) can be interpreted as a $(p+2)$-brane-antibrane pair
$(\mathcal{L},\Oh_Y)$ with the tachyon field $T$. Note that the
tachyon field is a holomorphic section of $L$ which vanishes along
$X$ and hence it has charges $(1,-1)$ under the U$(1)\times$U$(1)$
that live on the brane and antibrane. Here the U$(1)$ gauge field
on the brane is coming from the holomorphic connection on $L$,
with the same $p$-brane charge  as that of a $p$-brane wrapped on
$X$ which is associated to the restriction of the bundle $L$ to
$X$. Also, from the definition of the tachyon field $T$, we have
the trivial gauge field on the antibrane, with vanishing $p$-brane
charge. Now by taking the homology of the complex (2), we obtain
two sheaves $\mathcal{L}_X$, with $p$-brane charge and $\Oh_X$,
with $0$-brane charge, where $\mathcal{L}_X$ is the locally free
sheaf associated to the restriction bundle of $L$ to $X$. Thus we
see that the system is deformed to a system consisting of a
$p$-brane wrapped on $X$.
In the above discussion of brane-antibrane annihilation, we have
started from a rather strong condition for the choice of a
holomorphic line bundle on $Y$. With a suitable choice of a
tachyon field, we get a system of a brane-antibrane pair wrapping
$Y$, which is in vacuum state along $Y-X$. This is a basic
assumption made in \cite{witten98} for brane-antibrane
annihilation and the assumption  exactly agrees with that of
complexes giving a class in $K^0_{\mathsf{alg}}(Y,Y-X)$. The
duality between the relative $K$-group and $K$-homology,
discussed
in Section 3, shows that only such brane-antibrane systems on $Y$
can be deformed to a system of branes on $X$. In other words, a
brane-antibrane system on $Y$ in the absence of D-branes in $Y-X$
is deformed to a system of branes on $X$.

Now, let $X$ be a closed  subvariety of $Y$, which is of arbitrary
codimension.
For any complex of holomorphic bundles on $Y$ which are exact
off $X$,
it gives a brane-antibrane pair on $Y$ which is in vacuum state
along $Y-X$. By taking homology of the complex, we get a system of
branes on $X$.
In general, homology sheaves
are coherent sheaves. This suggests that we may not have a system of
single brane wrapped on $X$ when we deform a brane-antibrane
system on $Y$ to a system on $X$. On the other hand, it may be
possible to have a configuration consisting of several branes
wrapped on $X$ that is represented by a coherent sheaf. This
happens when $X$ is a singular variety.

A $p$-brane wrapped on $X$ also has in general lower-dimensional
brane charges. In order to fully describe all states with a
$p$-brane wrapped on $X$ in terms of states of a brane-antibrane
pair wrapped on $Y$, we need to choose a holomorphic line bundle
$K$ on $X$. Let $\mathcal{K}$ be the locally free sheaf associated
to the holomorphic bundle. Extension by zero gives a   coherent
sheaf $i_*\mathcal{K}$ on $Y$. This sheaf is resolved by locally
free sheaves on $Y$ and the associated complex of holomorphic
vector bundles on $Y$ is exact off $X$. Thus it defines a class
$K^0_{\mathsf{alg}}(Y,Y-X)$. The class associated to
$i_*\mathcal{K}$ is independent of the choice of a resolution
because any two such resolutions are dominated by the third. This
proves that the D-brane charge takes values in the relative
$K$-group when  a D$p$-brane is wrapped on $X$. The lower brane
charges are depend on the choice of a line bundle on $X$. This
shows that D-branes wrapping a nonsingular closed subvariety $X$
of a variety $Y$ are holomorphically classified by the relative
$K$-group $K^0_{\mathsf{alg}}(Y,Y-X)$. Since $Y$ is nonsingular,
the homology map, given in (1), is an isomorphism. Thus by
applying the homology map, we conclude that the D-brane charge is
properly classified by $K_0^{\mathsf{alg}}(X)$.

When $X$ is a singular variety,
there are no $p$-branes wrapping $X$
due to the anomaly condition in
\cite{freedwitten99}. However, there can be several
branes wrapping
subvarities of $X$ and hence
we can have a configuration
consisting of several different type of
branes wrapped on $X$, supporting suitable
holomorphic structures on each branes. On the whole,
the configuration must not be a holomorphic vector
bundle since its transition functions are not holomorphic around the
singular loci. Suppose we have a system that the
entire set of bound states is represented by a coherent sheaf.
Then we can apply the same argument as in the case of
nonsingular varieties. To be more precise, let $\mathcal M$ be
such a coherent sheaf. Since $Y$ is nonsingular,
the
sheaf $i_*\mathcal{M}$ is resolved by holomorphic vector bundles
on $Y$ and hence the associated complex of bundles defines a class
in $K^0_{\mathsf{alg}}(Y,Y-X)$. By applying the homology map, we
see that D-branes wrapping $X$ are classified by the $K$-homology
of $X$.

\section{Concluding remarks}

We have described the global version of bound state construction
in the algebraic geometry context. We showed how D-branes wrapping
$X$ are holomorphically
classified by the relative $K$-group
$K^0_{\mathsf{alg}}(Y,Y-X)$. By taking the homology map
discussed in Section 3, we conclude that D-branes wrapping $X$ are
properly classified by the $K$-homology group of $X$ including
some singular cases.

We can also make a  connection with topological $K$-homolgy theory
from our consideration. Since $X$ is a complex projective
algebraic variety, the underlying topological space of $X$ is a
finite simplicial complex. Thus we can study the topological
$K$-group and topological $K$-homology of $X$. As shown in
\cite{BaDoug84}, one has the map from the Grothendieck group of
coherent sheaves on $X$ to the topological $K$-homology of the
underlying topological space of $X$. However, in general, the map
is neither injective nor surjective, see examples in
\cite{sharpe99} and \cite{Har}. From our conclusion, we see that
D-branes wrapping $X$ take values in the topological $K$-homology
$X$. However, those groups do not give us the classification of
topological D-branes from our discussion, because the topological
$K$-homology and the Grothendieck group of coherent sheaves cannot
be identified.

\section*{Acknowledgements}
The author is indebted Prof. Hoil Kim for help and instructions in
algebraic geometry.
This work was supported by the
Brain Korea 21 Project in 2001.


\begin{thebibliography}{AAA}


\bibitem{witten98} Witten, E.: D-branes and $K$-theory,
{\it J. High Energy Phys. } {\bf 9812} (1998), 019; hep-th/9810188.

\bibitem{sen98} Sen, A.:
SO(32) spinors of type I and other solitons on brane-antibrane pair,
{\it J. High Energy Phys. } {\bf 9809} (1998), 023;
hep-th/9808141.


\bibitem{freedwitten99} Freed, D. S. and Witten, E.:
Anomalies in string theory with D-branes,
{\it Asian J. Math. } {\bf 3} (1999),  819-851;
hep-th/9907189.

\bibitem{per06} Periwal, V.:
D-brane charges and $K$-homology,
{\it J. High Energy Phys. } {\bf 0007} (2000), 041;
hep-th/0006223.

\bibitem{pol95} Polchinski, J.:
Dirichlet-branes and Ramond-Ramond charges,
{\it Phys. Rev. Lett. } {\bf 75} (1995) 4724-4727;
hep-th/9510017.

\bibitem{szabo01} Szabo, R. J.:  Superconnections, anomalies and
non-BPS brane charges,  hep-th/0108043.

\bibitem{AST} Asakawa, T., Sugimoto, S. and Terashima, S.:
D-branes, matrix theory and $K$-homology,
hep-th/0108085.

\bibitem{BaDoug84} Baum, P. and  Douglas, R. G.:
$K$ homology and index theory,
{\it Proc. Sympos. Pure Math.}  {\bf 38} Part I (1982),
117-173.

\bibitem{sharpe99} Sharpe, E.: D-branes, derived categories, and
Grothendieck groups,
{\it Nucler Phys. B} {\bf 561} (1999), 433-450;
hep-th/9902116.

\bibitem{BaFulMac79} Baum, P., Fulton, W. and MacPherson, R.:
Riemann-Roch and topological $K$-theory for singular
varieties,
{\it Acta Math.}  {\bf 143} (1979), 155-192.


\bibitem{Har} Hartshorne, R.:  {\it Algebraic Geometry}, Graduate
Texts in Mathematics  {\bf 52}, Springer-Verlag, 1977.

\bibitem{MoWi99} Moore, G. and Witten, E.:
Self-duality, Ramond-Ramond fields, and $K$-theory,
{\it J. High Energy Phys.} {\bf 0005} (2000), 032;
hep-th/9912279.

\bibitem{DMW} Diaconescu, D., Moore, G. and Witten, E.:
$E_8$ gauge theory, and a derivation of $K$-theory from
$M$-theory,
hep-th/0005090.


\bibitem{witten95} Witten, E.: Small instantons in string theory,
{\it Nucler Phys. B} {\bf 460} (1996), 541-559;
hep-th/9511030.

\bibitem{HarMoor}  Harvey, J. A. and Moore, G.:
On the algebras of BPS states,
{\it Comm. Math. Phys.} {\bf 197} (1998), 489-519;
hep-th/9609017.












\end{thebibliography}
\end{document}